\documentclass[12pt]{article}
\usepackage{authblk}
\usepackage{hyperref}
\usepackage[top=25truemm,bottom=28truemm,left=24truemm,right=24truemm]{geometry}
\usepackage{setspace}
\normalsize
\usepackage{amsmath,amssymb,mathrsfs,amsbsy,latexsym,amsfonts,amsthm}
\usepackage{fancyhdr}
\pdfoutput=1 
\usepackage{graphicx}
\usepackage{color}
\usepackage{comment}
\usepackage{here}
\usepackage{braket}
\usepackage[utf8]{inputenc}
\usepackage{mathtools}

\newcommand{\nn}{\nonumber\\}

\newcommand{\Z}{\mathbf{Z}}
\usepackage{epsf}

\numberwithin{equation}{section}
\title{\bf 
Wave Packets
in AdS/CFT Correspondence}

\author[1]{
	Seiji~Terashima\thanks{\tt terasima(at)yukawa.kyoto-u.ac.jp}
}
\vspace{5mm}

\affil[1]{\it\normalsize 
Center for Gravitational Physics and Quantum Information,  
\mbox{Yukawa Institute for Theoretical Physics, Kyoto University, Kyoto 606-8502, Japan}  }
\date{}

\begin{document}

\maketitle
\thispagestyle{fancy}
\renewcommand{\headrulewidth}{0pt}

\begin{abstract}

In this paper, we construct a general bulk wave packet
in the AdS/CFT correspondence.
This wave packet can be described both in bulk and CFT descriptions.
Then, we compute the time evolution of the energy density of this wave packet state on the vacuum in the CFT picture of $AdS_3/CFT_2$.
We find that the energy density of the wave packet is localized in two points, which means that
the bulk wave packet corresponds to two light-like particle-like objects
in the CFT picture.
Our result implies that 
the entanglement wedge reconstruction
given in \cite{Almheiri:2014lwa} is invalid.

\end{abstract}

\newpage
\thispagestyle{empty}
\setcounter{tocdepth}{2}

\setlength{\abovedisplayskip}{12pt}
\setlength{\belowdisplayskip}{12pt}

\tableofcontents
\newpage

\section{Introduction and summary}

The AdS/CFT correspondence \cite{Maldacena:1997re} is expected to be important to understanding quantum gravity.  
In particular, because the bulk space-time should emerge from CFT,
we can, in principle, understand the bulk space-time in quantum gravity from AdS/CFT correspondence.
For this purpose, several bulk space-time probes in CFT are known, including correlation functions, Wilson loops, and entanglement entropy.

The important probes of bulk space-time, which have not been studied intensively, are the wave packets in bulk space-time. 
Such wave packets are fundamental objects for thought experiments in bulk space-time.
In particular, a local region in the bulk space-time can be probed by the time-evolved wave packet. 
Then, the relationship between the local region of the bulk where the wave packet resides and the corresponding region of the boundary in the CFT picture will be the key to understanding how the bulk spacetime emerges from the CFT.
Thus, it is important to understand how the bulk wave packets are described in the CFT picture.
In \cite{Terashima:2021klf, Terashima:2020uqu}, the special kind of the bulk wave packets, for which only the direction of them is fixed, were considered in AdS/CFT correspondence, 
although the general bulk wave packets have not been studied.

In this paper, we first construct a general bulk wave packet
in the AdS/CFT correspondence.
Here, we take the large $N$ limit, which is the free limit in the bulk picture, and the generalized free approximation in the CFT picture.
Furthermore, we consider only the bulk scalar field and the corresponding scalar CFT operator, for simplicity.
This wave packet can be described both in bulk and CFT descriptions.\footnote{
In this paper, we consider the light-like wave packet. This is because we consider a generic $\Delta$ and the non-relativistic wave packet can be considered only for $\Delta \gg l_{AdS}$, where the $\Delta$ is the conformal dimension of the CFT operator and $l_{AdS}$ is the length scale of the AdS space, which we will set $l_{AdS}=1$.  
}

Then, we compute the energy density of this state of the wave packet on the vacuum in the CFT picture of $AdS_3/CFT_2$.\footnote{
More precisely, we will compute the energy density of this state of the wave packet represented by \eqref{wads} in the CFT picture.
The energy density depends on $1/N$ corrections
of the wave packet operator.
Our operator is special because it is localized on a point of the boundary on a time-slice.
}
Note that the energy density does not vanish if the CFT state 
is excited by the local CFT operator there.
Thus, if the distribution of the energy density is localized in some regions,
the CFT state is localized in that regions.
We find that the energy density of the wave packet is localized in two points, which are on the light-cone, in the CFT picture, which means that
the bulk wave packet corresponds to two light-like particle-like objects
in the CFT picture, although these are not like the free particles.
This result completely agrees with the result in \cite{Terashima:2021klf, Terashima:2020uqu}, 
in which the entanglement wedge reconstruction given in \cite{Almheiri:2014lwa} was shown to be invalid.
Note that our results in this paper only use the BDHM extrapolation relation \cite{Banks:1998dd}, which is the basic AdS/CFT dictionary, like the GKPW relation \cite{Gubser:1998bc, Witten:1998qj}, and the known three-point function in 2d CFT.

We also compute the VEV of the CFT primary scalar operator for the wave packet.
The distribution of this is completely different from the one of the energy density.
This is because there are infinitely many independent fields at a fixed time 
for the generalized free field.

\section{Wave packets in AdS/CFT}\label{sec:wave}

\subsection{Bulk and CFT fields in AdS/CFT}

Let us consider the global $AdS_{d+1}$ and
$\Omega$ represents coordinates of $(d-1)$-dimensional sphere $S^{d-1}$.
The coordinates $\tau$ and $\rho$ run in the range $-\infty<\tau<\infty$ and $0\leq \rho<\pi/2$.
In the coordinates, the metric takes
\begin{align}
    ds^2=\frac{1}{\cos^2 \rho}\left(-d\tau^2+d \rho^2+ \sin^2 \rho d\Omega_{d-1}^2\right).
\end{align}
For the Poincare patch of $AdS_{d+1}$,
the metric is 
\begin{align}
    ds^2=\frac{1}{z^2}\left(-dt^2+dz^2+ \delta_{i j} dx^i dx^j \right),
    \label{mp}
\end{align}
where $z>0$ and $i,j=1,2, \cdots, d-1$.

Let us consider the canonical quantization of free scalar field $\phi$ with mass $m$, which satisfies
the equations of motion, $(\Box-m^2) \phi=0$.
For the Poincare $AdS_{d+1}$, the mode expansion of the scalar field is given by the Bessel functions as 
\begin{align}
\phi(t,z,x^i) =
C \int_{\omega > \sqrt{k^2}} 
d \omega d k_i \,  e^{i \omega t-i k_j x^j } \, z^{\frac{d}{2}} \, a^\dagger_{\omega, k} J_{\nu}(\sqrt{ \omega^2 -k^2
} \, \, z) + h.c.,
\label{mode1}
\end{align}
where $\nu=\sqrt{m^2+d^2/4}= \Delta-d/2$ and
$k^2 \equiv  k^j k_j$.
Note that
we included the normalizable modes only.
Here, the asymptotic behavior of the Bessel function is
\begin{align}
 &  J_{\nu}(\sqrt{ \omega^2 -k^2
} \, \, z) \rightarrow \sqrt{ \omega^2 -k^2
}^\nu z^\nu \mbox{   for $z \rightarrow 0$}, \\
& J_{\nu}(\sqrt{ \omega^2 -k^2
} \, \, z) \rightarrow 
\sqrt{{2 \over \pi \sqrt{ \omega^2 -k^2
} z } } 
\cos (\sqrt{ \omega^2 -k^2
} z-\frac{2\nu+1}{4} \pi) \mbox{   for $z \rightarrow \infty$}.
\end{align}
The overall constant $C$ is usually chosen such that it satisfies the canonical commutator:
\begin{align}
    [\phi(t,z',{x'}^i), \frac{\partial}{\partial t} \phi(t,z,x^i) ]= i \delta (z'-z) \delta ({x'}^i-x^i),
\end{align}
where we defined the creation operators as
\begin{align}
    [a^\dagger_{\omega', k'}, a_{\omega, k} ]= \delta (\omega'-\omega) \delta ({k'}^i-k^i).
\end{align}
However, we will take a different choice as explained below.

The CFT primary operator $\cal O$ corresponding to the bulk scalar field $\phi$ is obtained by the BDHM relation ${\cal O}(t,x^i)= \lim_{z \rightarrow 0} \phi(t,z,x^i)/z^\Delta$ \cite{Banks:1998dd} as
\begin{align}
{\cal O} (t,x^i) =
C \int_{\omega > \sqrt{k^2}} 
d \omega d k_i \,  e^{i \omega t-i k_j x^j } \, 
\sqrt{ \omega^2 -k^2}^{\Delta-d/2} \,
a^\dagger_{\omega, k} 
+ h.c.,
\label{CFTp}
\end{align}
which is valid only for the large $N$ limit or the generalized free theory limit.
We will choose the normalization constant $C$ such that the above BDHM relation holds with 
the standard normalization of the 
CFT primary field ${\cal O} (t,x^i)$.
For $d=2$, this means 
\begin{align}
    &
    \bra{0}  {\cal O}( u_1, v_1) 
    {\cal O} ( u_2, v_2) \ket{0}
    = 
    \frac{1}{(u_1-u_2)^{ \Delta} (v_1-v_2)^{ \Delta} }
    .
\end{align}
where $u= (t+x), \, v=(t-x)$.

\subsection{Wave packets}

In this paper, we will consider essentially Gaussian wave packets because we will study the generic properties of the wave packets.
We will mainly consider one particle state on the bulk side for simplicity.
It is easy to generalize this state to the coherent state for the weak coupling bulk theory, as done in Appendix \ref{a1}.

\paragraph{Minkowski spacetime}

Let us remember the wave packets, at $t=\Vec{x}=0$,
of a free scalar field in $d+1$ dimensional Minkowski spacetime:
\begin{align}
    \int d \Vec{x} \, e^{-\frac{ \Vec{x}^2}{2 a^2} +i \Vec{p} \cdot \Vec{x}} \phi(t=0,\Vec{x}) \ket{0}
    \propto \int d \Vec{k} \, e^{-\frac{ a^2 (\Vec{k}-\Vec{p})^2}{2}} a^\dagger_{\Vec{k}} \ket{0},
    \label{g1}
\end{align}
where $\Vec{p}$ is the momentum of the wave packet
and
$[a^\dagger_{\Vec{k}}, a_{\Vec{k}}] =\delta (\Vec{k})$.
Instead of this, we can consider another wave packet
which is defined by Gaussian integral for the time and $x^i$ where $i=2, \cdots , d$:

\begin{align}
    \int d t \, \prod_{i=2,\cdots, d} dx^i \, e^{-\frac{  x^i x_i+t^2}{2 a^2}+i p_i x^i+i \omega t} \phi(t,\Vec{x})|_{x_1=0}\ket{0}
    \propto  \int d \Vec{k} \, 
    e^{- \frac{a^2}{2}  \left( (k^i-p^i)(k_i-p_i)+ (\sqrt{(k_1)^2+k^i k_i}-\omega)^2  \right) } 
    a^\dagger_{\Vec{k}} \ket{0},
     \label{g2}
\end{align}
where $i$ runs only for $2, \cdots, d$.
Here, we assume that 
 $|a p^i| \gg 1$ and $a \omega \gg 1$, which are needed for the wave packet and we always assume these. 
With these, it is approximated as
\begin{align}
    \int d \Vec{k} \, 
    e^{- \frac{a^2}{2}  
    \left( (\delta k^i)^2+ ( k_1-p^1 +\frac{p^i \delta k_i}{ p^1})^2 \right) } 
    a^\dagger_{\Vec{k}} \ket{0}
    +\int d \Vec{k} \, 
    e^{- \frac{a^2}{2}  
    \left( (\delta k^i)^2+ ( k_1+p^1 +\frac{p^i \delta k_i}{ p^1})^2 \right) } 
    a^\dagger_{\Vec{k}} \ket{0},
    \label{g3}
\end{align}
where $p^1=\sqrt{\omega^2-p^i p_i}$ and $\delta k^i=k^i-p^i$.
These are the Gaussian integrals around $k^1=\pm p^1$ and $k^i=p^i$.
Thus, the wave packet \eqref{g2} is essentially the same as the sum of the 
original wave packets \eqref{g1} with opposite momenta.
Note that if we change the Gaussian factor $e^{-\frac{  x^i x_i+t^2}{2 a^2}}$ in \eqref{g2} to a general one
$e^{-\frac{  h_{ij} x^i x^j+h'_{i} x^i t+ t^2}{2 a^2}}$ with the appropriate constants $h, h'$,
the approximated Gaussian factor can be taken to the same one as \eqref{g1}.
Furthermore, if we consider the theory on the half-space $x^1>0$ with a boundary condition on $x=0$, only one wave packet can be obtained by \eqref{g2}. 
We will consider such wave packets in $AdS_{d+1}$ where $x^1$ corresponds to the radial direction $z$.

\paragraph{Wave packet in AdS/CFT}

In AdS spacetime, wave packets are constructed as in flat spacetime.
The wave packets should be very small in size compared with the AdS scale, where
the AdS spacetime can be approximated as a Minkowski spacetime. 
Thus, the wave packets \eqref{g1} \eqref{g2} in the flat spacetime can be
regarded as the wave packets in AdS spacetime for $a \ll 1$.
Furthermore, in AdS spacetime, any wave packet will reach the asymptotic boundary by the time or backward time evolution.
Thus, we need to prepare wave packets almost on the boundary only to represent a general wave packet.
On the boundary, the bulk scalar field $\phi$ is identified as 
the CFT primary field ${\cal O}$ with an overall factor by the BDHM relation.
This means that the bulk wave packet in AdS/CFT can be given by
\begin{align}
    \ket{p,\bar{\omega}}
    &=\lim_{z \rightarrow 0} {1 \over z^\Delta}
    \int d t \, d x^i \, e^{-\frac{ x^i x_i+t^2}{2 a^2}+i p_i x^i-i \bar{\omega} t} 
\phi(t,z,x^i) \ket{0} \nn
    &=\int d t \, d x^i \, e^{-\frac{ x^i x_i+t^2}{2 a^2}+i p_i x^i-i \bar{\omega} t} {\cal O}(t,x) \ket{0},
     \label{wads}
\end{align}
for the Poincare $AdS_{d+1}$ \eqref{mp}.\footnote{
Using \eqref{CFTp}, we can rewrite the state in the momentum space:
\begin{align}
    \ket{p,\bar{\omega}}
    =
    (a \sqrt{\pi})^{d} C \int_{\omega > \sqrt{k^2}} 
d \omega d k_i \,  e^{-a^2 \frac{ (k_i-p_i)^2+(\omega-\bar{\omega})^2}{2}} \, 
\sqrt{ \omega^2 -k^2}^{\Delta-d/2} \,
a^\dagger_{\omega, k} 
    \ket{0},
     \label{wads2}
\end{align}
for the generalized free field approximation.
The norm of this state is 
\begin{align}
    {\cal N}^2=\langle p,\bar{\omega} \ket{p,\bar{\omega}}
    =
    (a \sqrt{\pi})^{2d} C^2 \int_{\omega > \sqrt{k^2}} 
d \omega d k_i \,  e^{-a^2 ((k_i-p_i)^2+(\omega-\bar{\omega})^2)} \, 
\sqrt{ \omega^2 -k^2}^{2 \Delta-d}, 
     \label{wadsn}
\end{align}
which is approximated as $ {\cal N}^2 \simeq (a \sqrt{\pi}^3)^{d} \sqrt{ \bar{\omega}^2 -p^2}^{2 \Delta-d} C^2$.
}
This can be regarded as the state in bulk and also the state in the CFT.
Here we require that 
\begin{align}
    a^2 p^2 \gg 1 , \,\,\,\, a \bar{\omega} \gg 1,
\end{align}
then the wave packet has the definite orientation with the momentum $p_i$ and energy $ \bar{\omega}$.

Let us check the time evolution of this state in the bulk picture.
The bulk localized (one-particle) state is
\begin{align}
\phi(t,z,x^i) \ket{0}=
C \int_{\omega > \sqrt{k^2}} 
d \omega d k_i \,  e^{i \omega t-i k_j x^j } \, z^{\frac{d}{2}} J_{\nu}(\sqrt{ \omega^2 -k^2 
} \, \, z) 
\, a^\dagger_{\omega, k}  \ket{0},
\label{bs}
\end{align}
which is not normalized. \footnote{This state can not be normalized and we need to smear this state to eliminate the high-energy modes. Here, we will use this because we will consider the overlap between this and the wave packet state which is already smeared.}
In order to consider the bulk spatial distribution of the wave packet state \eqref{wads} at time $t$,
we will consider the following overlap:
\begin{align}
& \bra{0} \phi(t=0,z,x^i) \,\,\, e^{i H t} \ket{p,\bar{\omega}} \nn
= &
 (a \sqrt{\pi})^{d} C^2 \int_{\omega > \sqrt{k^2}} 
d \omega d k_i \,  e^{i \omega t+i k_j x^j } e^{-\frac{a^2}{2} ((k_i-p_i)^2+(\omega-\bar{\omega})^2)} \, 
\sqrt{ \omega^2 -k^2}^{2 \Delta-d} \, z^{\frac{d}{2}} J_{\nu}(\sqrt{ \omega^2 -k^2 
} \, \, z).
\label{ov}
\end{align}
Because of the Gaussian factor, the integrals are dominated for the region near $k_i=p_i, \omega=\bar{\omega}$.
Defining $\delta \omega \equiv  \omega - \bar{\omega}, \,  \delta k_i \equiv  k_i -p_i$ 
and $p_z \equiv \sqrt{ \bar{\omega}^2 -p^2 } $,
The overlap can be approximated as
\begin{align}
& \bra{0} \phi(t=0,z,x^i) \,\,\, e^{i H t} \ket{p,\bar{\omega}} \nn
\sim &
 (a \sqrt{\pi})^{d} C^2 \sqrt{2/\pi} z^{\frac{d-1}{2}} (p_z)^{2 \Delta-d-1/2} \,e^{i \bar{\omega} t+i p_j x^j } \, 
 \nn 
\,\, & \times
 \int 
d \delta \omega d \delta k_i \,  e^{i \delta \omega t+i \delta k_j x^j } e^{-\frac{a^2}{2} ( (\delta k)^2+\delta \omega^2)}  \cos ({(p_z)^2+\bar{\omega} \delta \omega-p^i \delta k_i \over p_z} z-\frac{2\nu+1}{4} \pi).
\label{ov2}
\end{align}
The integrals in this is proportional to 
\begin{align}
 \int 
d \delta \omega d \delta k_i \,  e^{i \delta \omega (t\pm \bar\omega z/p_z)+i \delta k_j (x^j \mp p^j z/p_z) } e^{-\frac{a^2}{2} ( (\delta k)^2+\delta \omega^2)}
\simeq 
e^{-{ 1\over 2 a^2} \left( (t\pm \bar\omega z/p_z)^2+  (x^j \mp p^j z/p_z)^2 \right)},
\label{ov3}
\end{align}
which is strongly suppressed by the Gaussian factor for 
\begin{align}
|t\pm \bar\omega z/p_z| \gg a, \,\,\, \mbox{  or }
|x^j \mp p^j z/p_z| \gg a.
\end{align}
Thus, at each time $t>0$, 
the wave packet is localized at $z = { p_z \over \bar\omega} t, \,\, x^j =- {p^j 
 \over \bar\omega} t $ which is on the light-like trajectory from
the boundary point at $t=0$ with the energy $\bar\omega$ and the momentum $p_z, p_i$, as expected.
This also implies that the size of the wave packet is ${\cal O}(a)$ for any time $t$ in the coordinate $z, x^i$.

\paragraph{Remarks on the asymptotic AdS case}

So far, we have considered the wave packets in AdS space.
The wave packet state \eqref{wads} can be regarded as the 
wave packet state in the asymptotic AdS case also.
This is because the state is written by the bulk field on the asymptotic boundary or 
the CFT primary fields.
Indeed, near the boundary, space-time can be regarded as the AdS space and
the state will represent the wave packet moving toward the inside of the asymptotic AdS space.
The wave packet will be on the null geodesics of the asymptotic AdS space.

Another remark is that the wave packet state \eqref{wads} can be created from the vacuum or the semi-classical background by the 
source term in CFT because 
it is written by the CFT primary fields.
Indeed, by adding the source term $ e^{\int dt' dx' \epsilon J(t',x') {\cal O} (t',x') }$ with small $\epsilon$,
we can obtain the wave packet state \eqref{wads} in the sub-leading order in $\epsilon$
by setting the source term $J(t,x)$ as the Gaussian factor in \eqref{wads}.
For general $\epsilon$, the state becomes the coherent state given in Appendix \ref{a1}.
In the bulk theory, the one-particle state is described by the free quantum field approximation around 
the AdS background and the coherent state can be considered as a free approximation of the classical field.

\section{Energy density of wave packet in CFT picture }

We can consider the time-evolution of the bulk wave packet \eqref{wads} as a state in CFT. 
Here, the bulk wave packet can be regarded as a basic probe of the bulk spacetime point.
Thus, in order to understand how the bulk spacetime emerges from CFT, 
it is important to know what is the spacetime region of this state in the CFT picture.\footnote{ 
For the special kind of the bulk wave packet and the corresponding CFT state was 
considered in \cite{Terashima:2021klf, Terashima:2020uqu} and we will see that the general bulk wave packet \eqref{wads}
also have the same property, as expected.
}

We will compute the time-evolution of the energy density of the bulk wave packet \eqref{wads} in the CFT picture, in order to investigate where the state \eqref{wads} in the CFT picture is localized at each time.
Another quantity that might behave like the energy density is 
the expectation value of the primary scalar operators $\cal O$.
However, this quantity is not good for our purpose. 
Indeed, if we regard it represents the location of the state in the CFT picture, its time evolution violates causality, as we will see later.
The reason for this is as follows.
The generalized free field does not obey equations of motion
and $\frac{\partial^n}{\partial t^n}{\cal O}$ with different $n$ are independent at each time.
The expectation values of $\frac{\partial^n}{\partial t^n}{\cal O}$ are independent quantities. 
Thus, if these are different, we can not take one of them as a representative. 
Furthermore, because the number of these independent operators is infinite, 
it is difficult to obtain information on the location of the state in the CFT picture.

The energy density for \eqref{wads} is the three-point function of the two primary scalar operators and the energy-momentum tensor:
\begin{align}
   & \bra{p,\bar\omega}   T_{00}(t=\bar{t}, x^i=\bar{x}^i) \ket{p,\bar\omega} \nn
   =
   & \int d t_1 \, d x_1^i \, e^{-\frac{ (x_1^i)^2+t_1^2}{2 a^2}-i p_i x_1^i+i \bar{\omega} t_1} 
   \int d t_2 \, d x_2^i \, e^{-\frac{ (x_2^i)^2+t_2^2}{2 a^2}+i p_i x_2^i-i \bar{\omega} t_2}  \nn
   & \times \bra{0} {\cal O}(t_1,x_1) T_{00}(t=\bar{t}, x^i=\bar{x}^i)
   {\cal O}(t_2,x_2) \ket{0},
   \label{ed1}
\end{align}
in the Heisenberg picture.
Because such a three-point function in CFT is known exactly,
we can compute the energy density.
It is important to note that this computation does not use the generalized free approximation, which is the leading order of the large $N$ expansion, and 
the result is valid for a large, but finite $N$.

We also note that operator ordering of this is not the time ordering. 
The ordering of the operators is fixed by the path of analytic continuation from the Euclidean correlation function. This can be implemented by slightly deforming the insertion points by a small imaginary time, as in the $i \epsilon$ prescription.  
For \eqref{ed1}, 
we change $t_1 \rightarrow t_1 +i \epsilon_1$,  $\bar{t} \rightarrow \bar{t} +i \epsilon_T$ and $t_2 \rightarrow t_2 +i \epsilon_2$,
where $\epsilon_1 > \epsilon_T > \epsilon_2$.
For example, we can take $\epsilon_1 =-\epsilon_2 =\epsilon > 0 $ and $\epsilon_T=0$.

Below, we will consider the energy density in the approximation $a \ll 1$ and $\bar\omega a \gg 1, (p_i)^2 a^2 \gg 1$.
We will neglect the terms which become zero in this limit. 
We also assume $\Delta ={\cal O} (1)$,
which implies the mass $m$ is ${\cal O} (1)$
and 
the wave packet behaves like a massless particle because the energy and the momentum of it are much larger than the mass.\footnote{
If we take $\Delta \gg 1$ such that 
$m a \ll 1$, the Compton length $1/m$ will be larger than the size of the wave packet.
In this case, the wave packet can correspond to the massive particle. 
}
We will also consider the $d=2 $ case only. 

\subsection{$d=2$}

For $CFT_2$ in the complex plane, the energy-momentum tensor is given by $T(z)$ and $\bar{T}(\bar{z})$.
We need to compute $\bra{0}  {\cal O}(z',\bar{z}') T_{00}(\xi,\bar{\xi}) {\cal O}(z,\bar{z}) \ket{0}$
where the energy density is $\frac{1}{2 \pi} T_{00}(z,\bar{z})=\frac{1}{2 \pi} (T(z)+\bar{T}(\bar{z}))$.
Using the conformal Ward identity, 
the three-point function\footnote{
Here, we consider Euclidean, not Lorentzian, CFT and we do not care about the operator ordering as usual
although the operator ordering is fixed by the imaginary time. 
}
is evaluated (see, for example, \cite{Belin:2019mnx}) as
\begin{align}
     \bra{0} T (\xi) {\cal O}_1 (z_1,\bar{z}_1) {\cal O}_2 (z_2,\bar{z}_2) \ket{0}
    &=
    \sum_{i=1,2} 
    \left( \frac{h_i}{(\xi-z_i)^2} +\frac{\partial_i}{\xi-z_i} \right)
    \bra{0} {\cal O}_1 (z_1,\bar{z}_1) {\cal O}_2 (z_2,\bar{z}_2) \ket{0} \nn
    &= \sum_{i=1,2} 
    \left( \frac{h_i}{(\xi-z_i)^2} +\frac{\partial_i}{\xi-z_i} \right)
    \frac{1}{(z_1-z_2)^{\Delta_1+\Delta_2} (\bar{z}_1-\bar{z}_2)^{\Delta_1+\Delta_2} },
\end{align}
where we used the normalization of the primary operator such that the two-point function is the standard one and
$h_i$ is the weight with $\Delta=h+\bar{h}$.

For Minkowski spacetime,
we will replace $z \rightarrow  u= (t+x), \, \bar{z} \rightarrow -v=-(t-x)$,
then we obtain\footnote{
More precisely, we need to include the small imaginary part as the $\epsilon$-prescription
according to the ordering of the operators.
The definitions of $\frac{1}{(u_1-u_2)^{ \Delta}}$ and similar terms should be 
given by the convergent sum expression in the Euclidean cylinder as explained in, for example, \cite{Nagano:2021tbu}.
In our case, although the overall phase factor depends on these definitions,
it is indeed fixed by requiring that the energy is real and non-negative.
} 
\begin{align}
    & \bra{0} {\cal O}(t_1,x_1) T_{00}(t=\bar{t}, x=\bar{x})
   {\cal O}(t_2,x_2) \ket{0}
   =
    \bra{0}  {\cal O}( u_1, v_1) 
    ( T ( \bar{u}) +\bar{T} ( \bar{v}))
    {\cal O} ( u_2, v_2) \ket{0} \nn
    = & 
      \sum_{i=1,2} 
    \left( \frac{\Delta/2}{(\bar{u}-u_i)^2} +\frac{\partial_{u_i}}{\bar{u}-u_i} + 
    \frac{\Delta/2}{(\bar{v}-v_i)^2} +\frac{\partial_{v_i}}{\bar{v}-v_i} \right)
    \frac{1}{(u_1-u_2)^{ \Delta} (v_2-v_1)^{ \Delta} } \nn
    = & 
      \frac{\Delta}{2} 
     \left(
      \frac{(u_1-u_2)^2}{(\bar{u}-u_1)^2(\bar{u}-u_2)^2}
      +
        \frac{(v_1-v_2)^2}{(\bar{v}-v_1)^2(\bar{v}-v_2)^2}
      \right)
    \frac{1}{(u_1-u_2)^{ \Delta} (v_2-v_1)^{ \Delta} }
    .
\end{align}
Using this, the energy density for \eqref{wads} is 
\begin{align}
   & \bra{p,\bar\omega}   T_{00}(t=\bar{t}, x=\bar{x}) \ket{p,\bar\omega} \nn
   =
   & \int d t_1 \, d x_1 \, e^{-\frac{ (x_1)^2+t_1^2}{2 a^2}-i p x_1+i \bar{\omega} t_1} 
   \int d t_2 \, d x_2^i \, e^{-\frac{ (x_2)^2+t_2^2}{2 a^2}+i p x_2-i \bar{\omega} t_2}  \nn
   & \times \frac{\Delta}{2} 
     \left(
      \frac{(u_1-u_2)^2}{(\bar{u}-u_1)^2(\bar{u}-u_2)^2}
      +
        \frac{(v_1-v_2)^2}{(\bar{v}-v_1)^2(\bar{v}-v_2)^2}
      \right)
    \frac{1}{(u_1-u_2)^{ \Delta} (v_2-v_1)^{ \Delta} }.
\end{align}

Below we will evaluate this explicitly.
Because the calculation is not very technically simple, 
we will state the results of the calculation first.
The energy density ${\cal E}(t,x)$ of the wave packet state is approximately given by
\begin{align}
   {\cal E}(t,x)
      \simeq & 
  \frac{1}{2 \sqrt{2 \pi} a}
  \left( e^{-\frac{ (x+t)^2}{2 a^2}  } (\bar\omega -p) +e^{-\frac{ (x-t)^2}{2 a^2}  } (\bar\omega+p)) \right)
    .
\end{align}
which is localized on the light cone $x=t$ 
or $x=-t$.

Before computing the energy density of the wave packet state, 
let us consider the state $\ket{p,\bar\omega} $ with $p=\bar\omega=0$ because it is simpler.
This corresponds to the (Gaussian smeared) local CFT operator insertion.
We will see that the energy density of this state in CFT is localized on the light cone.
For this, we have
\begin{align}
   & \bra{0,0}   T_{00}(t=\bar{t}, x=\bar{x}) \ket{0,0} \nn
   =
   & \int d t_1 \, d x_1
   \int d t_2 \, d x_2  \, e^{-\frac{ (x_1)^2+(t_1)^2}{2 a^2}} \, e^{-\frac{ (x_2)^2+(t_2)^2}{2 a^2}}  \nn
   & \times  \frac{\Delta}{2} 
     \left(
      \frac{(u_1-u_2)^2}{(\bar{u}-u_1)^2(\bar{u}-u_2)^2}
      +
        \frac{(v_1-v_2)^2}{(\bar{v}-v_1)^2(\bar{v}-v_2)^2}
      \right)
    \frac{1}{(u_1-u_2)^{ \Delta} (v_2-v_1)^{ \Delta} }.
\end{align}
The Gaussian integral approximately vanish 
except for the region near $u_i=v_i=0$.
For $|\bar{u}| \gg a$ and $|\bar{v}| \gg a$, which implies the energy-momentum tensor is inserted far from the light cone of the scalar operator insertion point,
we can use the following expansion:
\begin{align}
   & \bra{0,0}   T_{00}(t=\bar{t}, x=\bar{x}) \ket{0,0} \nn
   = &
   \int d t_1 \, d x_1
   d t_2 \, d x_2  \, e^{-\frac{ (x_1)^2+t_1^2+ (x_2)^2+t_2^2}{2 a^2}}  
   \frac{\Delta}{2} 
     \left(
      \frac{(u_1-u_2)^2}{\bar{u}^4}
      +
        \frac{(v_1-v_2)^2}{\bar{v}^4}
      \right)
    \frac{1}{(u_1-u_2)^{ \Delta} (v_2-v_1)^{ \Delta} }
    +\cdots,
\end{align}
where $\cdots$ means ${\cal O}(\frac{1}{\bar{u}^5})$ and $ {\cal O}(\frac{1}{\bar{v}^5})$ terms.
This implies that 
$\bra{0,0}   T_{00}(t=\bar{t}, x^i=\bar{x}^i) \ket{0,0} \sim {\cal O}(\frac{1}{|\bar{x}\pm \bar{t}|^4}) $
at $t=\bar{t}$
and the contribution of the region
$|\bar{x}\pm \bar{t}|\gg a$ to the energy of the state at $t=\bar{t}$
is ${\cal O}(\frac{1}{|\bar{x}\pm \bar{t}|^3}) $.
Thus, the energy density is localized and only non-zero, for $a \rightarrow 0$, at $\bar{x}= \bar{t}$ or $\bar{x}= -\bar{t}$, which are on the light cone.

Now, let us go back to studying the wave packet state $\ket{p,\bar\omega} $.
For this, 
by defining 
\begin{align}
    p_u= \bar\omega -p \,\, (\geq 0), \,\, p_v = \bar\omega +p \,\, (\geq 0),
\end{align}
we have
\begin{align}
   & \bra{p,\bar\omega}   T_{00}(t=\bar{t}, x=\bar{x}) \ket{p,\bar\omega} \nn
   =
   & \int d t_1 \, d x_1 \, e^{-\frac{ (x_1)^2+t_1^2}{2 a^2}-i p x_1+i \bar{\omega} t_1} 
   \int d t_2 \, d x_2 \, e^{-\frac{ (x_2)^2+t_2^2}{2 a^2}+i p x_2-i \bar{\omega} t_2}  \nn
   & \times \frac{\Delta}{2} 
     \left(
      \frac{(u_1-u_2)^2}{(\bar{u}-u_1)^2(\bar{u}-u_2)^2}
      +
        \frac{(v_1-v_2)^2}{(\bar{v}-v_1)^2(\bar{v}-v_2)^2}
      \right)
    \frac{1}{(u_1-u_2)^{ \Delta} (v_2-v_1)^{ \Delta} } \nn
    =
    & \frac{1}{4}\int d u_1 \, d v_1 \, d u_2 \, d v_2 \,\, e^{-\frac{ (u_1)^2+(v_1)^2+(u_2)^2+(v_2)^2}{4 a^2}+i (p_u u_1+p_v v_1-p_u u_2-p_v v_2)/2}
     \nn
   & \times \frac{\Delta}{2} 
     \left(
      \frac{(u_1-u_2)^2}{(\bar{u}-u_1)^2(\bar{u}-u_2)^2}
      +
        \frac{(v_1-v_2)^2}{(\bar{v}-v_1)^2(\bar{v}-v_2)^2}
      \right)
    \frac{1}{(u_1-u_2)^{ \Delta} (v_2-v_1)^{ \Delta} } \nn
    =
    & \frac{1}{4}\int d u_1 \, d v_1 \, d u_2 \, d v_2 \,\, e^{-\frac{ (u_1-i p_u a^2)^2+(v_1-i p_v a^2)^2+(u_2+i p_u a^2)^2+(v_2+i p_v a^2)^2}{4 a^2} 
    - \frac{a^2}{2} ((p_u)^2+(p_v)^2) }
     \nn
   & \times \frac{\Delta}{2} 
     \left(
      \frac{(u_1-u_2)^2}{(\bar{u}-u_1)^2(\bar{u}-u_2)^2}
      +
        \frac{(v_1-v_2)^2}{(\bar{v}-v_1)^2(\bar{v}-v_2)^2}
      \right)
    \frac{1}{(u_1-u_2)^{ \Delta} (v_2-v_1)^{ \Delta} },
\end{align}
which is localized on the light cone $\bar{u}=0$ 
or $\bar{v}=0$, as we can easily see using the same argument above.
We will evaluate this more explicitly below.
Let us consider a part of it:
\begin{align}
    A \equiv & \int d u_1 \, d v_1 \, d u_2 \, d v_2 \,\, e^{-\frac{ (u_1-i p_u a^2)^2+(v_1-i p_v a^2)^2+(u_2+i p_u a^2)^2+(v_2+i p_v a^2)^2}{4 a^2} 
    - \frac{a^2}{2} ((p_u)^2+(p_v)^2) }
     \nn
   & \times  
      \frac{(u_1-u_2)^2}{(\bar{u}-u_1)^2(\bar{u}-u_2)^2}
    \frac{1}{(u_1-u_2)^{ \Delta} (v_2-v_1)^{ \Delta} } \nn
    = & 
    e^{-\frac{a^2}{2} ((p_v)^2) } 
    \int  d v_1  \, d v_2 \,\, e^{-\frac{ (v_1-i p_v a^2)^2+(v_2+i p_v a^2)^2}{4 a^2} 
    } 
    \frac{1}{(v_2-v_1)^{ \Delta}}
    \nn
    & \times
    e^{-\frac{a^2}{2} ((p_u)^2 } 
    \int d u_1 \,  d u_2  \,\, e^{-\frac{ (u_1-i p_u a^2)^2+(u_2+i p_u a^2)^2}{4 a^2}  }
    \frac{1}{(\bar{u}-u_1)^2(\bar{u}-u_2)^2}
    \frac{1}{(u_1-u_2)^{ \Delta-2}  }
    .
\end{align}
The other part is obtained by interchanging 
$\{ \bar{u} , p_u \} $ and $\{ \bar{v} , p_v \}$.
Here, the integration paths are taken as $u_i \in \mathbf{R} - i \epsilon_i$ 
and $v_i \in \mathbf{R} - i \epsilon_i$ with $\epsilon_2 > 0 > \epsilon_1$ for the $i \epsilon$-prescription for the ordering of the operator.
First, we will perform $v_1$-integration  
\begin{align}
   & e^{-\frac{a^2}{2} ((p_v)^2) } 
    \int  d v_1  \, d v_2 \,\, e^{-\frac{ (v_1-i p_v a^2)^2+(v_2+i p_v a^2)^2}{4 a^2} 
    } 
    \frac{1}{(v_2-v_1)^{ \Delta}} \nn
    =& (-1)^{\Delta} e^{-\frac{a^2}{2} ((p_v)^2) } 
    \int  d v_1  \, d v_2 \,\,  
    \frac{1}{(v_1-v_2)^{ \Delta-q}}
    \frac{\Gamma(\Delta-q)}{\Gamma(\Delta)}
    \frac{\partial^q}{\partial v_1^q} \, e^{-\frac{ (v_1-i p_v a^2)^2+(v_2+i p_v a^2)^2}{4 a^2} 
    }
    ,
    \label{no1}
\end{align}
where $q$ is the integer such that $1 \geq \Delta-q>0$, 
and consider what happens if we move the path to 
$v_1 \in \mathbf{R} + i a^2 p_v$. 
Then, the integration does not depend on $p_v$ and
the Gaussian factor $e^{-\frac{a^2}{2} (p_v)^2}$, which is very small, can not be canceled. 
Thus, this contribution can be neglected in our approximation.
The remaining parts of this are contributions from the pole or the branching point at $v_1=v_2$ in the region between   
$\mathbf{R} - i \epsilon_1$ and $ \mathbf{R} + i a^2 p_v$.

For $\Delta \in \mathbf{Z}$, for which $q=\Delta-1$,
the contribution from the single pole at $v_1=v_2$ is
\begin{align}
  & (-1)^{\Delta} \frac{2 \pi i }{\Gamma(\Delta)}
  e^{-\frac{a^2}{2} ((p_v)^2) } 
    \int d v_2 \,\,  
    \left.
    \frac{\partial^{\Delta-1}}{\partial v_1^{\Delta-1}} \, e^{-\frac{ (v_1-i p_v a^2)^2+(v_2+i p_v a^2)^2}{4 a^2} 
    } \right|_{v_1 =v_2} \nn
    \simeq & 
    \frac{2 \pi i }{\Gamma(\Delta)}
  ( -i p_v/2)^{\Delta-1} \,
  e^{-\frac{a^2}{2} ((p_v)^2) } 
    \int d v_2 \,\,  
     e^{-\frac{ (v_2)^2-(p_v a^2)^2}{2 a^2} 
    } \nn
    = & 
    \frac{(2 \pi)^{3/2}  }{\Gamma(\Delta)} (-i)^{\Delta}
  a ( p_v/2)^{\Delta-1} \,
    ,
    \label{di}
\end{align}
where we have neglected the terms which are sub-leading in $1/(p_v a)$ expansion.\footnote{ 
After the Gaussian $v_2$-integration, $v_2$ becomes ${\cal O}(a)$ quantity.
}

For $\Delta \notin \mathbf{Z}$, there is a branching point $v_1=v_2$.
Here, we can neglect the contribution around the branching point
because $1 > \Delta-q $.
Thus, the contribution from the cut from $v_1=v_2$ is
\begin{align}
   & (-1)^{\Delta} e^{-\frac{a^2}{2} ((p_v)^2) } 
    \int  d v_2  \, 
    (1-e^{2 \pi i (q-\Delta)})
    i \int_{0}^{p_v a^2}  dy \,\,  
    \frac{1}{(i y)^{ \Delta-q}}
    \frac{\Gamma(\Delta-q)}{\Gamma(\Delta)}
    \frac{\partial^q}{\partial (i y)^q} \, e^{-\frac{ (v_2+iy-i p_v a^2)^2+(v_2+i p_v a^2)^2}{4 a^2} 
    } 
    ,
\end{align}
where $v_1=v_2+ i y$.
Because of the Gaussian factor, which is almost zero for $y \gg a$, the $ y$ integration can be approximated as
\begin{align}
   & 
    \int_{0}^{p_v a^2}  dy \,\,  
    \frac{(-1)^{\Delta}}{(i y)^{ \Delta-q}}
    \frac{\partial^q}{\partial (i y)^q} \, e^{-\frac{ (v_2+iy-i p_v a^2)^2}{4 a^2} 
    } 
    \simeq 
    \int_{0}^{\infty}  dy \,\,  
    \frac{(-1)^{\Delta}}{(i y)^{ \Delta-q}}
    (i p_v/2)^q \, e^{-y p_v /2 }  e^{-\frac{ (v_2-i p_v a^2)^2}{4 a^2} } \nn
    = & \Gamma(1-\Delta+q) ( p_v/2)^{\Delta-1} e^{-\frac{ (v_2-i p_v a^2)^2}{4 a^2}}(-1)^{\Delta-q} (-i)^{\Delta}
    .
\end{align}
Then, we can easily check that this gives \eqref{di}
for $\Delta \notin \mathbf{Z}$ also.

Next, we will compute
\begin{align}
  & e^{-\frac{a^2}{2} ((p_u)^2 } 
    \int d u_1 \,  d u_2  \,\, e^{-\frac{ (u_1-i p_u a^2)^2+(u_2+i p_u a^2)^2}{4 a^2}  }
    \frac{1}{(\bar{u}-u_1)^2(\bar{u}-u_2)^2}
    \frac{1}{(u_1-u_2)^{ \Delta-2}  } \nn
    =& e^{-\frac{a^2}{2} ((p_u)^2 } 
    \int  d u_2  \,\, e^{-\frac{ (u_2+i p_u a^2)^2}{4 a^2}  }
    \frac{1}{(\bar{u}-u_2)^2}
    \int d u_1 \,
    \frac{1}{(u_1-\bar{u})}
    \frac{\partial}{\partial u_1} \left(
     e^{-\frac{ (u_1-i p_u a^2)^2}{4 a^2}  }
    \frac{1}{(u_1-u_2)^{ \Delta-2}  } \right)
    .
    \label{u1u2}
\end{align}
As for the $v_1$ integration, 
we move the path of the $u_1$ integration to 
$u_1 \in \mathbf{R} + i a^2 p_u$ and 
take the residue at $u_1=\bar{u}$.\footnote{
There is the contribution from the singularities at $u_1 =u_2$ also.
However, at this, the Gaussian factor becomes
$e^{-\frac{ (u_2+i p_u a^2)^2}{4 a^2}  } e^{-\frac{ (u_2-i p_u a^2)^2}{4 a^2}  }=e^{-\frac{ (u_2)^2- (p_u a^2)^2}{2 a^2}  }$
and the $u_2$ integration is $p_u$ independent.
Using this, we can easily see that this contribution is smaller than the one from the pole at $u_1=\bar{u}$.
}
Then, the result is 
\begin{align}
 & 2 \pi i e^{-\frac{a^2}{2} ((p_u)^2 } e^{-\frac{ (\bar{u}-i p_u a^2)^2}{4 a^2}  }
    \int  d u_2  \,\, e^{-\frac{ (u_2+i p_u a^2)^2}{4 a^2}  }
    \frac{1}{(\bar{u}-u_2)^\Delta}
     \left(  i p_u/2-(\Delta-2) 
     \frac{1}{(\bar{u}-u_2)}
    \right)
    ,
\end{align}
where we have neglected the term proportional to $\bar{u}$,
which is small because there is the Gaussian factor $e^{-\frac{ (\bar{u})^2}{2 a^2}  }$ 
after the $u_2$ integration,
as we will see below. 
For the $u_2$ integration,  
we move the path to 
$u_2 \in \mathbf{R} - i a^2 p_u$ and 
take the residue at $u_2=\bar{u}$.
Then, the result is 
\begin{align}
 &  -(-1)^\Delta (2 \pi)^2  e^{-\frac{ (\bar{u})^2}{2 a^2}  }
 \frac{1}{\Gamma(\Delta)}
      (-i p_u/2)^{\Delta}
     \left( -1+
     \frac{\Delta-2}{\Delta}
    \right)
    =
      (2 \pi)^2  e^{-\frac{ (\bar{u})^2}{2 a^2}  }
 \frac{1}{\Gamma(\Delta)}
      (i p_u/2)^{\Delta}
     \frac{2}{\Delta}
    .
\end{align}
Thus, we obtain
\begin{align} 
    A \simeq \, & e^{-\frac{ (\bar{u})^2}{2 a^2}  }
      (2 \pi)^{5/2}  
 \frac{1}{\Gamma(\Delta)^2}
      (p_v p_u/4)^{\Delta-1}
     \frac{p_u a}{\Delta}
    .
\end{align}
We need to compute the normalization of the state: 
\begin{align}
    {\cal N}^2= &\bra{p,\bar\omega}   p,\bar\omega \rangle \nn
   =
   & \int d t_1 \, d x_1 \, e^{-\frac{ (x_1)^2+t_1^2}{2 a^2}-i p x_1+i \bar{\omega} t_1} 
   \int d t_2 \, d x_2 \, e^{-\frac{ (x_2)^2+t_2^2}{2 a^2}+i p x_2-i \bar{\omega} t_2} 
    \frac{1}{(u_1-u_2)^{ \Delta} (v_2-v_1)^{ \Delta} } \nn
    =
    & \frac{1}{4}\int d u_1 \, d v_1 \, d u_2 \, d v_2 \,\, e^{-\frac{ (u_1-i p_u a^2)^2+(v_1-i p_v a^2)^2+(u_2+i p_u a^2)^2+(v_2+i p_v a^2)^2}{4 a^2} 
    - \frac{a^2}{2} ((p_u)^2+(p_v)^2) }
     \nn
   & \times 
    \frac{1}{(u_1-u_2)^{ \Delta} (v_2-v_1)^{ \Delta} },
\end{align}
which is the same as the computation 
of \eqref{no1} and \eqref{di}.
Hence, we obtain
\begin{align}
    {\cal N}^2 \simeq
    \frac14 \,\,
     \frac{(2 \pi)^{3/2}  }{\Gamma(\Delta)} (-i)^{\Delta}
  a ( p_v/2)^{\Delta-1} \times 
   \frac{(2 \pi)^{3/2}  }{\Gamma(\Delta)} (i)^{\Delta}
  a ( p_u/2)^{\Delta-1}. 
\end{align}

Finally, we have the energy density ${\cal E}(\bar{t},\bar{x})$ of the wave packet state as 
\begin{align}
   {\cal E}(\bar{t},\bar{x})= &  \frac{1}{{\cal N}^2} \, \bra{p,\bar\omega} \frac{1}{2 \pi}  T_{00}(t=\bar{t}, x=\bar{x}) \ket{p,\bar\omega} \nn
      \simeq & 
  \frac{1}{2 \sqrt{2 \pi} a}
  \left( e^{-\frac{ (\bar{u})^2}{2 a^2}  } p_u +e^{-\frac{ (\bar{v})^2}{2 a^2}  } p_v \right)
    .
\end{align}
which is localized on the light cone $\bar{u}=0$ 
or $\bar{v}=0$.
The energies of the regions near $\bar{u}=0$ and $\bar{v}=0$ are $p_u/2$ and $p_v/2$, respectively,
because $\frac{1}{ \sqrt{2 \pi} a} \int dx
   e^{-\frac{ x^2}{2 a^2}  }=1$.
The sum of them is the energy $\bar\omega$of the state  correctly.

In summary, the energy density of the wave packet state \eqref{wads} in $AdS_3/CFT_2$ at time $t$
is localized on the light cone $x=\pm t $ for small $a$ and small $ 1/( \bar\omega a)$.
The energy localized near $x= t$, which is equivalent to $v=x-t =0$, is 
$(\bar\omega +p)/2$. 
The energy localized near $x= -t$, which is equivalent to $u=x+t =0$, is 
$(\bar\omega -p)/2$. 
Thus, the wave packet on the bulk corresponds to a pair of the excitations at $t > 0$,
which is given schematically by $(\tilde{\cal O}_v(x=t) + \tilde{\cal O}_u(x=-t) )\ket{0}$
where $\tilde{\cal O}_v, \tilde{\cal O}_u$ are some local operators.
For $\bar\omega =p$, the state is only at $x=t$, not a pair.
Indeed, in the bulk picture also, the wave packet is on the boundary for $\bar\omega =p$,
then it is localized at $x=t, z=0$ in the bulk picture.

\paragraph{Global AdS/CFT}

\begin{figure}[htbp]
  \begin{minipage}[b]{0.45\linewidth}
    \centering
    \includegraphics[width=10cm]{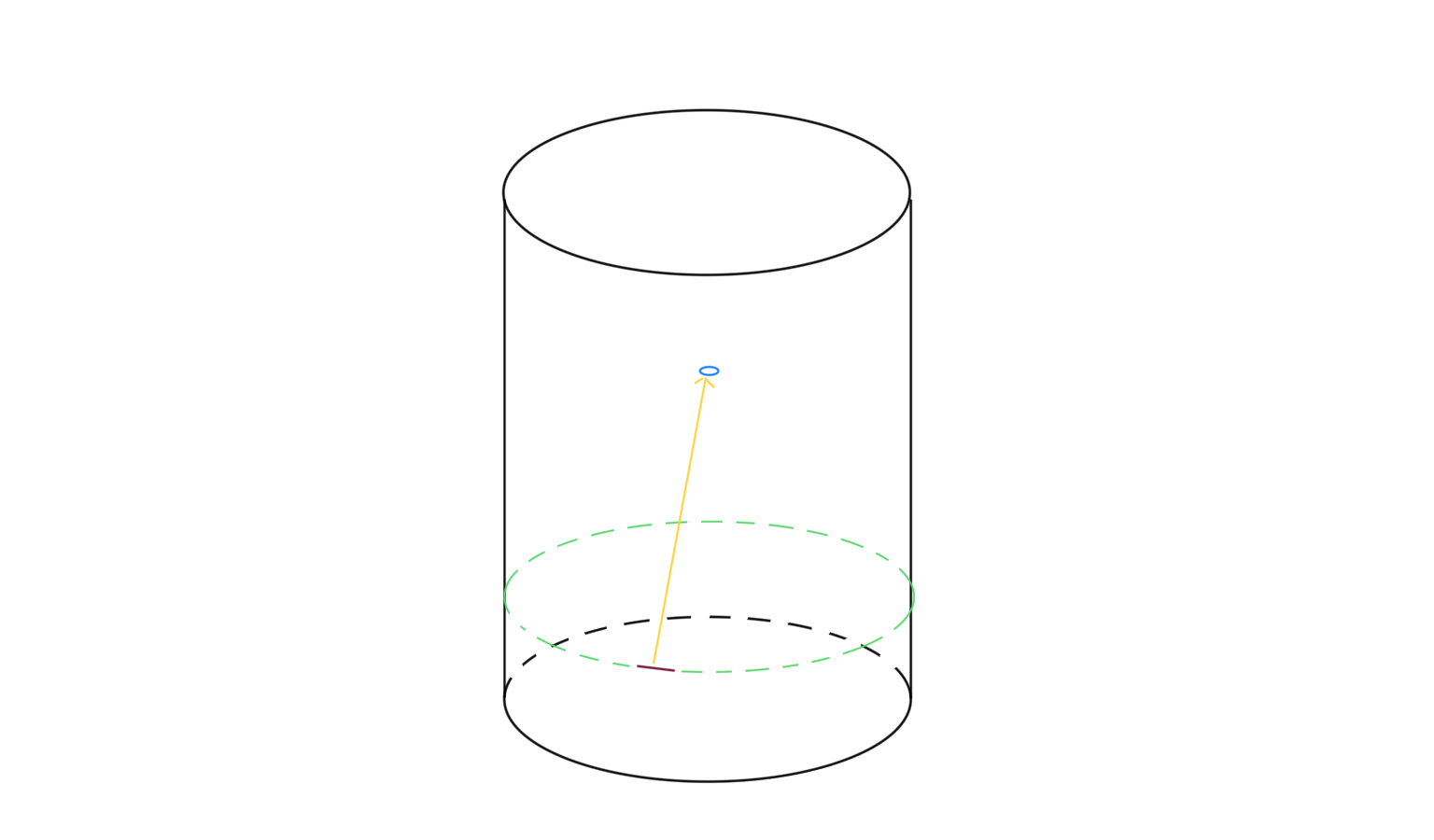}
    \caption{An example of the bulk wave packet (moving toward the center).}
    \label{f1}
  \end{minipage}
  \begin{minipage}[b]{0.45\linewidth}
    \centering
    \includegraphics[width=10cm]{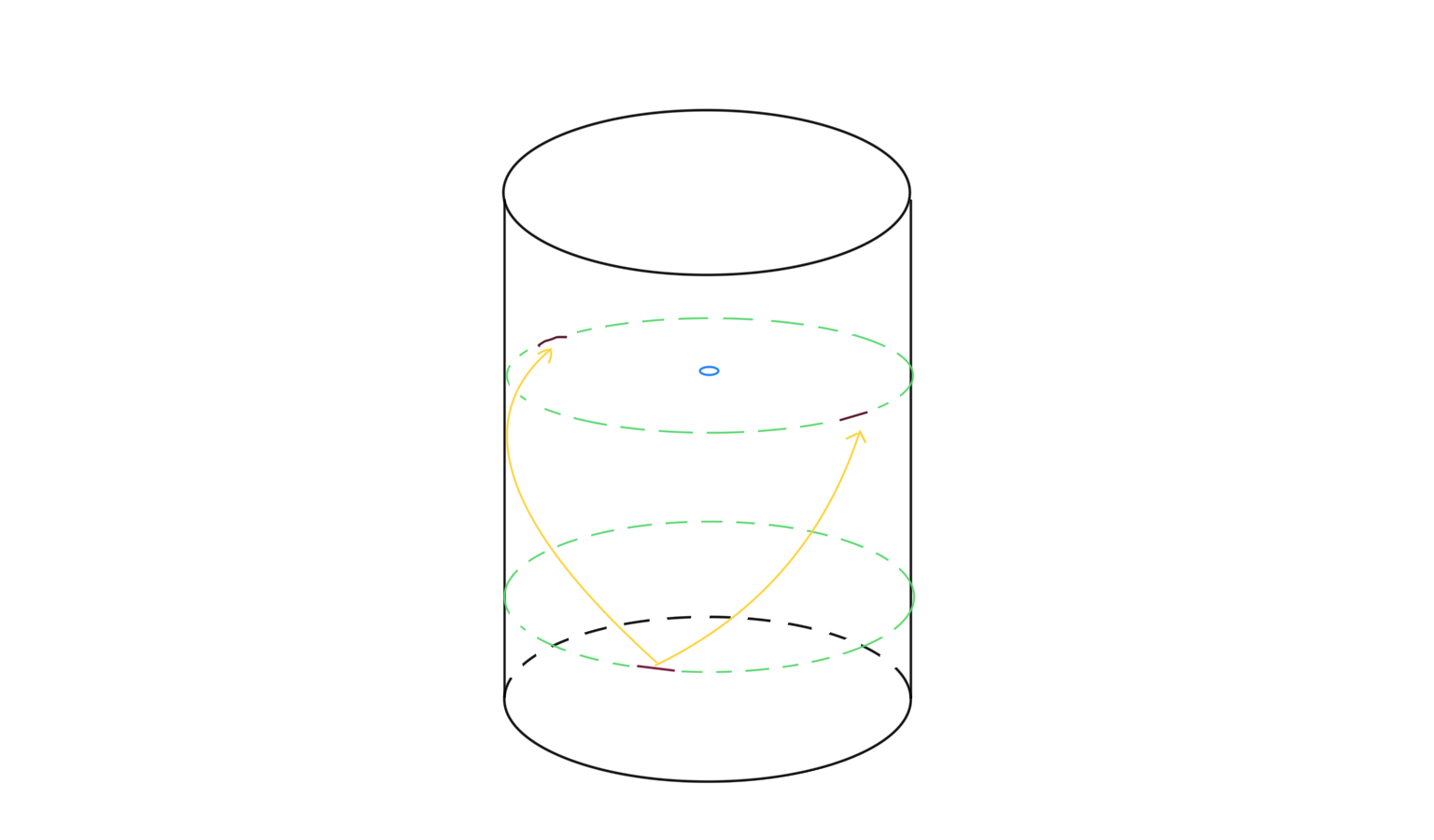}
    \caption{The corresponding two "particles" in the CFT picture}
    \label{f2}
  \end{minipage}
\end{figure}

So far, we have considered the wave packets in the Poincare AdS case.
We can easily generalize this to the global AdS case by the conformal transformation.
(More precisely, for the global $AdS_3/CFT_2$ case,
the parameters in the wave packet state \eqref{wads} should be replaced by, for example,
$z \rightarrow \pi/2-\rho$, $t \rightarrow \tau $ and $x \rightarrow \tanh (\theta)$,
where $ -\pi<\theta \le \pi$ is the coordinate for the $S^1$.)
Instead of explicitly doing this, we can conclude that 
the above summary for the Poincare AdS case is also true for the  global AdS case.
This is because the computations of the energy density essentially use
the information on the singularities of the three-point function. 
Thus, the CFT picture of the bulk wave packet by the energy density 
is the same as the "simple bulk reconstruction" picture given in \cite{Terashima:2021klf}.
See Fig. \ref{f1} and Fig. \ref{f2}.
Note that only this is consistent with the causalities in both the bulk and the boundary theories because the bulk wave packet starting from the boundary at $t=0$ will reach the boundary at $t=\pi$, and then the bulk local field can be regarded as the CFT local primary field at $t=0$ and $t=\pi$.   

One might think that our result is inconsistent with the HKLL bulk reconstruction formula \cite{Hamilton:2006az} because the bulk local state corresponds to only two points on the light cone.
However, as shown in \cite{Terashima:2021klf}
this is consistent with the HKLL bulk reconstruction formula because of the ambiguities of the smearing function in the formula.
In Appendix \ref{b2}, 
we will summarize the discussion for it.

\paragraph{Sub-region duality and entanglement wedge reconstruction}

If the sub-region duality and the entanglement wedge reconstruction are correct,
if we take a region $A$ in the CFT picture such that  
the bulk wave packet is in the bulk entanglement $M_A$ wedge for $A$,
the state should be supported only in the region $A$ in the CFT picture.\footnote{
Here, the sub-region duality and the entanglement wedge reconstruction means those given in \cite{Almheiri:2014lwa}, in which the bulk Hilbert space was assumed to be a tensor product of the two Hilbert spaces for the subregion $M_A$ and $M_{\bar{A}}$.
This may be (approximately) realized by a gauge fixing,
for example, the FG gauge.
In particular, we claim that the global and Rindler HKLL bulk reconstructions of a bulk local operator in the overlap of the two entanglement wedges should be different in the leading order of the $1/N$-expansion.
}
This is not the case if the wave packet is the horizon to the horizon type discussed in \cite{Terashima:2021klf}. 
Here, this statement takes into account the energy density from the perspective of CFT, which is one of the leading effects of bulk interactions.
This invalidity of sub-region duality and the entanglement wedge reconstruction\footnote{
This invalidity may be due to the invalidity of the large $N$ expansion for the AdS/CFT correspondence for the subregion \cite{Sugishita:2022ldv}.
This is related to the brick wall in AdS/CFT \cite{Iizuka:2013kma} \cite{tHooft:1984kcu}
and the fuzz ball conjecture \cite{Mathur:2005zp} \cite{Mathur:2009hf}.
}
can be 
seen from a simpler example.
Let us consider a state that is obtained by acting the (smeared) bulk local operator $\phi$ at the center of the global AdS space, i.e. $\rho=0$, on the vacuum. 
This can be written by the CFT operator by the HKLL bulk reconstruction formula \cite{Hamilton:2006az}.
Then, the energy density is obtained by an explicit calculation.
Of course, by the symmetry of state, the result is a uniform distribution on $S^{d-1}$.
Then, if we take a region $A$ in the CFT picture such that 
the bulk wave packet is in the bulk entanglement wedge $M_A$, 
the state should be supported only in the region $A$ in the CFT picture according to
the subregion duality and the entanglement wedge reconstruction.
This means that there exists
an operator supported in the subregion $A$ that produces the non-zero energy density
outside $A$. It is obviously unphysical.

Here, it should be stressed that 
there exists CFT operator ${\cal O}_A$, supported in the region $A$, corresponding to the bulk operator $\phi_{M_A}$, supported in the region $M_A$ such that
${\cal O}_A \ket{\psi} = \phi_{M_A} \ket{\psi}$
where $\ket{\psi}$ is an arbitrary low energy state
if the entanglement wedge reconstruction is correct.
This implies that
\begin{align}
   {\cal O}^2_A {\cal O}^1_A \ket{\psi} =  \phi^2_{M_A}  \phi^1_{M_A} \ket{\psi}
   \label{ew}
\end{align}
 for $ {\cal O}^i_A \ket{\psi} = \phi^i_{M_A} \ket{\psi}$
because by writing $\ket{\psi_1}=\phi^1_{M_A} \ket{\psi}$ we find 
${\cal O}^2_A \ket{\psi_1} = \phi^2_{M_A} \ket{\psi_1}$.
Then, the Reeh-Schlieder theorem (and the mirror map of the thermofield double) is not useful for the entanglement wedge reconstruction
although they can give a similar CFT operator such that ${\cal O}_A \ket{0} = \phi_{M_A} \ket{0}$
which does not satisfy \eqref{ew}.
Furthermore, with the Reeh-Schlieder theorem, the vacuum acting by operators supported in any subregion can give any state. 
Thus, any small subregion can be dual to the whole space, and statements of the subregion duality and entanglement wedge reconstruction will be meaningless using the Reeh-Schlieder theorem.

Below, we will show that the entanglement wedge reconstruction is violated for the coherent state of the wave packet explicitly.  
First, we define  
\begin{align}
   \phi_{p,\bar\omega}=\lim_{z \rightarrow 0} {1 \over z^\Delta} \int d t \, d x^i \, e^{-\frac{ x^i x_i+t^2}{2 a^2} } 
   (e^{i p_i x^i-i \bar{\omega} t} +e^{-i p_i x^i+i \bar{\omega} t})
   \phi(t,z,x^i),
     \label{xxxxxx}
\end{align}
and consider $\ket{\psi}=e^{i \epsilon  \phi_{p,\bar\omega}} \ket{0}$  represented by the CFT primary operator like \eqref{wads}.
Note that $ \ket{p,\bar{\omega}} \simeq \phi_{p,\bar\omega} \ket{0}$ for $\bar\omega \gg 1/a \gg 1$.
We can show that $\bra{\psi} T_{00}(t,x) \ket{\psi}= \bra{p,\bar{\omega}} T_{00} (t,x)\ket{p,\bar{\omega}} +{\cal O}(\epsilon^3)$ because 
$\bra{0} T_{00} (t,x)  \phi_{p,\bar\omega}^2
 \ket{0} \simeq 0$, which follows from the fact that the pole at $u_1=0$ or $u_2=0$
 does not contribute to the $u_1, u_2$ integration  in \eqref{u1u2} for this ordering of the operators.
Let us choose $p,\bar{\omega}$ such that  at $t=\bar{t}$  the bulk  wave packet is in $M_A$
and the energy density in the CFT picture is nonzero at some points in $A$ and in $\bar{A}$.
On the other hand, if 
$ \phi_{p,\bar\omega}$ is reconstructed from the CFT operator on $A$, i.e.
$ \phi_{p,\bar\omega}= \phi_{p,\bar\omega} ({\cal O}_A )$,
we can show $\bra{\psi} T_{00}(t=\bar{t}, x=\bar{x}) \ket{\psi}=0$ for $x \in \bar{A}$
because 
\begin{align}
\bra{\psi} T_{00}(t=\bar{t}, x=\bar{x}) \ket{\psi}=
    \bra{0} ([T_{00}(t=\bar{t}, x=\bar{x}) , i \epsilon \phi_{p,\bar\omega} ({\cal O}_A )]+ \cdots
    )
    \ket{0}=0,
\end{align}
because $A$ and $t=\bar{t}, x=\bar{x}$ are spatially separated and the causality of the CFT implies the commutators are zero.
Thus, $\phi_{p,\bar\omega} $, which is supported on $M_A$, can not be reconstructed from the CFT operator supported on $A$.

\subsection{Overlap between the wave packet state and CFT local state}

Instead of the energy density, one might think the 
overlap between the wave packet state $\ket{p,\bar\omega}$ and CFT local state ${\cal O} (x,t) \ket{0}$
will give some information about where the state is localized or the spatial distribution of the state.
This can be essentially regarded as 
the VEV of the scalar operator for the coherent state, which is discussed in the Appendix,
because $\bra{J} {\cal O} (x,t) \ket{J} \rightarrow \epsilon \bra{0} (\int dt' dx' \epsilon J(t',x') {\cal O} (t',x') \, {\cal O} (x,t)  +  {\cal O} (x,t) \int dt' dx' \epsilon J(t',x') )\ket{0} = \epsilon (\bra{p,\bar\omega} {\cal O} (x,t) \ket{0} + \bra{0} {\cal O} (x,t)  \ket{p,\bar\omega})   $ for small $\epsilon$, where
$\ket{J}= e^{\int dt' dx' \epsilon J(t',x') {\cal O} (t',x')} \ket{0}$ is the coherent state.\footnote{
Even for the case that $\epsilon$ is not small, the similar expression holds because the 
${\cal O} (x,t)$ is linear in the creation and annihilation operators in the generalized free approximation.
}
The VEV of the scalar operator is one of the most important calculable quantities in AdS/CFT, at least if it is time-independent.
Nevertheless, 
it is highly difficult to obtain any information on the properties of this state in CFT.
This is because 
$ \frac{\partial^n}{\partial t^n}{\cal O} (x,t) \ket{0}$
with any $n \ge 0$ is an independent state for fixed $x,t$ in the generalized free approximation, which is the large $N$ limit we have taken.
This means that there are infinitely independent states at each point in the large $N$ limit.
Thus, even if we know some information on $ {\cal O} (x,t)$ at fixed $t$,
it is a piece of infinitesimal information.\footnote{
For the energy-momentum tensor, the statements here can be applied.
However, the energy-momentum tensor has special properties.
The energy density will be non-negative and any local excitation will give a non-zero energy density.
Thus, the energy density can be used to understand the spatial distribution of the wave packet state.
}
With $ {\cal O} (x,t)$ for all $t$,
we can construct $ \frac{\partial^n}{\partial t^n}{\cal O} (x,t)$.
However, in order to do so, we need to know the $n$-th derivative coefficient precisely for arbitrary $n$.\footnote{
Even for time-dependent cases,
it is possible to obtain some information from
$ {\cal O} (x,t)$ depending on the state,
for example, the state that represents a wave in AdS/BCFT \cite{Izumi:2022opi}.
}

If we know information on $ {\cal O} (x,t)$ for any $x,t$,
we could recover, for example, the spatial distribution of the wave packet, in principle.
However, this should be non-local in time.
Furthermore, it is unclear how to recover it.
Indeed, there may be ambiguities for it as like the computation of the mutual information for
the generalized free field discussed in \cite{Benedetti:2022aiw}.
Therefore, we conclude that 
it is highly difficult to obtain any information on the properties of the wave packet state in CFT
using the overlap between it and the CFT local state or the VEV of the CFT operator.
Below, we will explicitly see the difficulty for the global $AdS_3$ case.

Let us consider the global $AdS_3$.
The CFT primary field in the large $N$ limit is written by
the creation and annihilation operators \cite{Terashima:2017gmc} as
\begin{align}
{\cal O} (\tau,\theta) \sim
 \sum_{n \in \Z_{\ge 0}, \, m \in \Z} 
\,  e^{i (2n+|m|) \tau-i m \theta} \, 
a^\dagger_{n,m} 
+ h.c.,
\label{gCFTp}
\end{align}
where we took $\Delta=d/2$ for simplicity.
Note that this is invariant under $\tau \rightarrow \tau+\pi$ and  $\theta \rightarrow \theta +\pi$.
We can easily extend this for general $\Delta$.
The bulk wave packet state at $\tau=0, \theta=0$ with the energy $\omega$ and the momentum $ p$ is given by
\begin{align}
\ket{p, \bar\omega}=\int d \tau d \theta 
\, e^{-\frac{ \tau^2+\theta^2}{2 a^2}  +i p \theta-i \bar{\omega} \tau}
{\cal O} (\tau,\theta) \ket{0} \sim
 a^2 \sum_{n \in \Z_{\ge 0}, \, m \in \Z} 
\,  e^{ -\frac{a^2}{2} ( ((2n+|m|) -\bar\omega)^2 + (m-p)^2 )} \, 
a^\dagger_{n,m} \ket{0},
\label{gCFTp2}
\end{align}
where we used the Gaussian factor $ e^{-\frac{ \theta^2}{2 a^2}}$ instead of $ e^{-\frac{ \tan (\theta)^2}{2 a^2}}$ because their difference is negligible for $a \ll 1$.
Then, the overlap is 
\begin{align}
\bra{0} {\cal O} (\tau,\theta) \ket{p, \bar\omega} 
& \sim
 a^2 \sum_{n \in \Z_{\ge 0}, \, m \in \Z} 
\,  e^{ -\frac{a^2}{2} ( ((2n+|m|) -\bar\omega)^2 + (m-p)^2 )} \, 
 e^{-i (2n+|m|) \tau+i m \theta}, \nn
 & \simeq
 a^2 e^{-i \bar\omega \tau+i p \theta}
 \sum_{n \in \Z, \, m \in \Z} 
\,  e^{ -\frac{a^2}{2} ( ((2n+|m|) )^2 + (m)^2 )} \, 
 e^{-i (2n+m) \tau+i m \theta}, \nn
 & \simeq a^4 e^{-i \bar\omega \tau+i p \theta} \delta(\tau+ \pi \Z)  \delta(\theta-\tau+ 2 \pi \Z)  
\label{gCFTp3}
\end{align}
where we assumed $p >0$ other than $|p| \gg 1, \bar\omega \gg 1$
and we noted, for example, the Gaussian factor $ e^{-\frac{\tau^2}{2 a^2}} /a^2$ as $\delta(\tau)$, for notational simplicity.
Thus, the distribution of it is localized on $\theta=0$  at $\tau=0 + 2 \pi \Z$ and
$\theta=\pi$  at $\tau=\pi + 2 \pi \Z$. These space-time points are when the bulk wave packet is at the boundary.
For other $\tau$, it almost vanishes. This means that 
the state is diffused to infinitely many states $ \frac{\partial^n}{\partial t^n}{\cal O} (x,t) \ket{0}$
for it.


\section*{Acknowledgement}

The author would like to thank N. Tanahashi for collaboration in the early stages of this work
and helpful discussions.
The author thanks S. Sugishita for the useful comments.
This work was supported by JSPS KAKENHI Grant Number 17K05414.
This work was supported by MEXT-JSPS Grant-in-Aid for Transformative Research Areas (A) ``Extreme Universe'', No. 21H05184.

\hspace{1cm}

Note added:
As this paper was being completed, we became aware of the preprint \cite{Kinoshita:2023hgc} in which
the bulk wave packet similar to ours was constructed and its VEV of the CFT operator was discussed.

\appendix

\section{Coherent state}
\label{a1}

We will denote the bulk local operator in the free limit as
\begin{align}
    \phi(X)=\sum_n \psi_n(X) a_n+h.c. \, .
\end{align}
where $X$ represents the all coordinates of the bulk space-time, $n$ labels the mode expansion
and $a_n$ represents the annihilation operator.
The corresponding CFT primary operator in the limit is given as
\begin{align}
    {\cal O}(x)=\sum_n \psi^{CFT}_n(x) a_n+h.c. \, .
\end{align}
where $x$ represents the all coordinates of the CFT space-time.
Then, the (normalized) semi-classical state in the limit is given by the coherent state:
\begin{align}
   \ket{\alpha} = e^{\sum_n (\alpha_n a^\dagger_n-\alpha^*_n a_n)} \ket{0}=e^{-\frac12 \sum_n |\alpha_n|^2} e^{\sum_n \alpha_n a^\dagger_n} \ket{0},
\end{align}
for which the time evolution is given by
\begin{align}
 \ket{\alpha(t)}=  e^{i H t} \ket{\alpha} = e^{\sum_n (e^{i E_n t} \alpha_n a^\dagger_n-e^{-i E_n t} \alpha^*_n a_n)}.
\end{align}
The VEV of the bulk and CFT local operators, which are linear in the creation and annihilation operators, are given as
\begin{align}
    \bra{\alpha} \phi(X) \ket{\alpha} =\sum_n \psi_n(X) \, \alpha_n+c.c. \, ,\,\,\, \nn
    \bra{\alpha} {\cal O}(x) \ket{\alpha} =\sum_n \psi^{CFT}_n(x) \, \alpha_n+c.c. \, .
\end{align}

Let us rewrite the one-particle state for the bulk wave packet \eqref{wads} as
\begin{align}
    \ket{p,\bar{\omega}}
    &=(\phi^{wp}) \ket{0},
     \label{wadsApp}
\end{align}
where 
\begin{align}
  \phi^{wp}=\sum_n \psi^{wp}_n a_n^\dagger = \int d t \, d x^i \, e^{-\frac{ x^i x_i+t^2}{2 a^2}
  +i p_i x^i-i \bar{\omega} t} {\cal O}^+ (t,x), 
\end{align}
where ${\cal O}^+ (t,x)$ is a part of ${\cal O} (t,x)$ which is linear in $a^\dagger_n$
and 
\begin{align}
  \psi^{wp}_n = \int d t \, d x^i \, e^{-\frac{ x^i x_i+t^2}{2 a^2}
  +i p_i x^i-i \bar{\omega} t} ( \psi^{CFT}_n(t,x^i) )^* . 
\end{align}
Note that the overlaps between this and the bulk and CFT local states considered in this paper are  
given by
\begin{align}
    &\bra{0} \phi(X)  \ket{p,\bar{\omega}} =\sum_n (\psi_n (X))^* \, \psi^{wp}_n \nn
    &\bra{0} {\cal O}(x)   \ket{p,\bar{\omega}} =\sum_n (\psi^{CFT}_n (X))^* \, \psi^{wp}_n.
\end{align}
Then, the corresponding coherent state representing the bulk wave packet is 
given by setting $\alpha_n =(\psi^{wp}_n)^*$, i.e.
\begin{align}
   \ket{wp} = e^{\sum_n ((\psi^{wp})^* a^\dagger_n-\psi^{wp}_n a_n)} \ket{0}.
\end{align}
The VEVs of the bulk and CFT local operators for this state are given as
\begin{align}
    &\bra{wp} \phi(X) \ket{wp} =\sum_n \psi_n(X) \,(\psi^{wp}_n)^* +c.c. 
    =\bra{0} \phi(X)  \ket{p,\bar{\omega}}  +c.c.
    \, ,\,\,\,  \nn
    &\bra{wp} {\cal O}(x) \ket{wp} =\sum_n \psi^{CFT}_n(x) \, (\psi^{wp}_n)^*+c.c.
    =\bra{0} {\cal O}(x) \ket{p,\bar{\omega}} + c.c. \, ,
\end{align}
which means that where the overlaps are distributed for the one particle state is the same as where the VEVs are distributed for the coherent state.
Thus, the corresponding coherent state represents the wave packet.

\section{On HKLL bulk reconstruction formula}
\label{b2}

The HKLL bulk reconstruction formula \cite{Hamilton:2006az} is the formula representing the bulk local field as the space-time integrals of the corresponding CFT primary operators in the generalized free limit, following the ideas in 
\cite{Banks:1998dd} \cite{Bena:1999jv} \cite{Duetsch:2002hc}.
The explicit formula for the bulk local field at the center in global AdS space is given by
\begin{align} 
\phi(\rho=0,\tau=0)
=\int_{-\frac{\pi}{2} \leq \tau \leq \frac{\pi}{2} } d \tau' d \Omega' K( \Omega' , \tau') {\cal O}(\Omega', \tau'),
\label{hkll}
\end{align}
where 
\begin{align} 
K(\Omega , \tau)  \sim 
{1 \over | \cos \tau |^{d-\Delta} }.
\end{align}
First, we note that 
the CFT primary operator ${\cal O}(\Omega, \tau)$ is periodic, i.e. ${\cal O}(\Omega, \tau+ 2 \pi)={\cal O}(\Omega, \tau)$,
in the generalized free field approximation.
We also note that there are 
infinitely many choices of 
the smearing function $K(\Omega , \tau)$
as noted in \cite{Hamilton:2006az}
because the Fourier transformation of the CFT primary operator ${\cal O}(\Omega, \tau)$ 
in the generalized free field approximation.
Indeed, they use this ambiguity to obtain this simple formula.

In \cite{Terashima:2021klf, Terashima:2020uqu},
the dominant contributions in the $\tau$ integrals in \eqref{hkll} are from $\tau=\pm \pi/2$.
This means, for example, that 
the overlap $\bra{\phi} {\cal O},\tau \rangle$ between the bulk local state
$\ket{\phi}=\phi(\rho=0,\tau=0)\ket{0}$
and the spherical symmetric CFT state $\ket{{\cal O},\tau}=\int d \Omega' {\cal O}(\Omega', \tau)\ket{0}$ is zero
for $\tau=\pi/2 + \pi \Z$.
Note that 
the smearing function can be essentially taken to 
$K(\Omega , \tau) \sim \bra{\phi} {\cal O},\tau \rangle $, which is zero
for $\tau \neq \tau=\pi/2 + \pi \Z$ 
because 
for the generalized free field we can show that $\bra{{\cal O},\tau} {\cal O},\tau' \rangle \sim \delta(\tau-\tau'+ \pi \Z)$.

The above statements are not precise because the bulk local operator and the CFT operator are ill-defined by the UV divergences.
More precisely, we can show that 
$\bra{\phi;a} {\cal O},\tau; a \rangle \rightarrow 0$ for $a \rightarrow 0$.
Here,
the smeared bulk local state at the center $\ket{\phi;a}$ is given by 
\begin{align}
\ket{\phi;a}=\frac{1}{{\cal N_\phi}^2}\int d\tau' e^{-\frac{(\tau')^2}{2 a^2}}\phi(\rho=0,\tau')\ket{0},     
\end{align}
where ${\cal N_\phi}^2$ is fixed by the normalization condition $\bra{\phi,a} \phi,a \rangle=1$,
and 
the smeared spherical symmetric CFT state $\ket{{\cal O},\tau;a}$ is given by 
\begin{align}
\ket{{\cal O},\tau;a}=\frac{1}{{\cal N_O}^2}\int d\tau' e^{-\frac{(\tau'-\tau)^2}{2 a^2}}{\cal O}(\tau')\ket{0},     
\end{align}
where ${\cal N_O}^2$ is fixed by the normalization condition $\bra{{\cal O},\tau,a} {\cal O},\tau,a \rangle=1$.
The smearing by the Gaussian integral roughly corresponds to the energy cut-off with $1/a$.

This seems to be impossible, in particular for $\tau=\pm \pi/2$ because
$K(\Omega , \tau=\pm \pi/2)=0$.
However, it is indeed possible because of the choice or the ambiguity of the smearing function.
The important point here is that the 
smearing function should be a periodic function, then it is singular at $\tau=\pi/2 + \pi \Z$ because of the absolute value of the $|\cos(\tau)|$.
The local field contains an arbitrarily high-energy mode and the singularity gives the non-trivial contribution to the arbitrarily high-energy mode.
Thus, the singular points give the dominant contribution to the reconstruction of the bulk local operator,
as shown in \cite{Terashima:2021klf}.\footnote{
The bulk local operator can be explicitly written by the CFT local operators only on $\tau=\pi/2$ with the time derivative \cite{Terashima:2021klf}
using the formulas given in 
\cite{Terashima:2017gmc}
\cite{Terashima:2019wed}
}

\begin{figure}[htbp]
  \begin{minipage}[b]{0.45\linewidth}
    \centering
    \includegraphics[width=7.5cm]{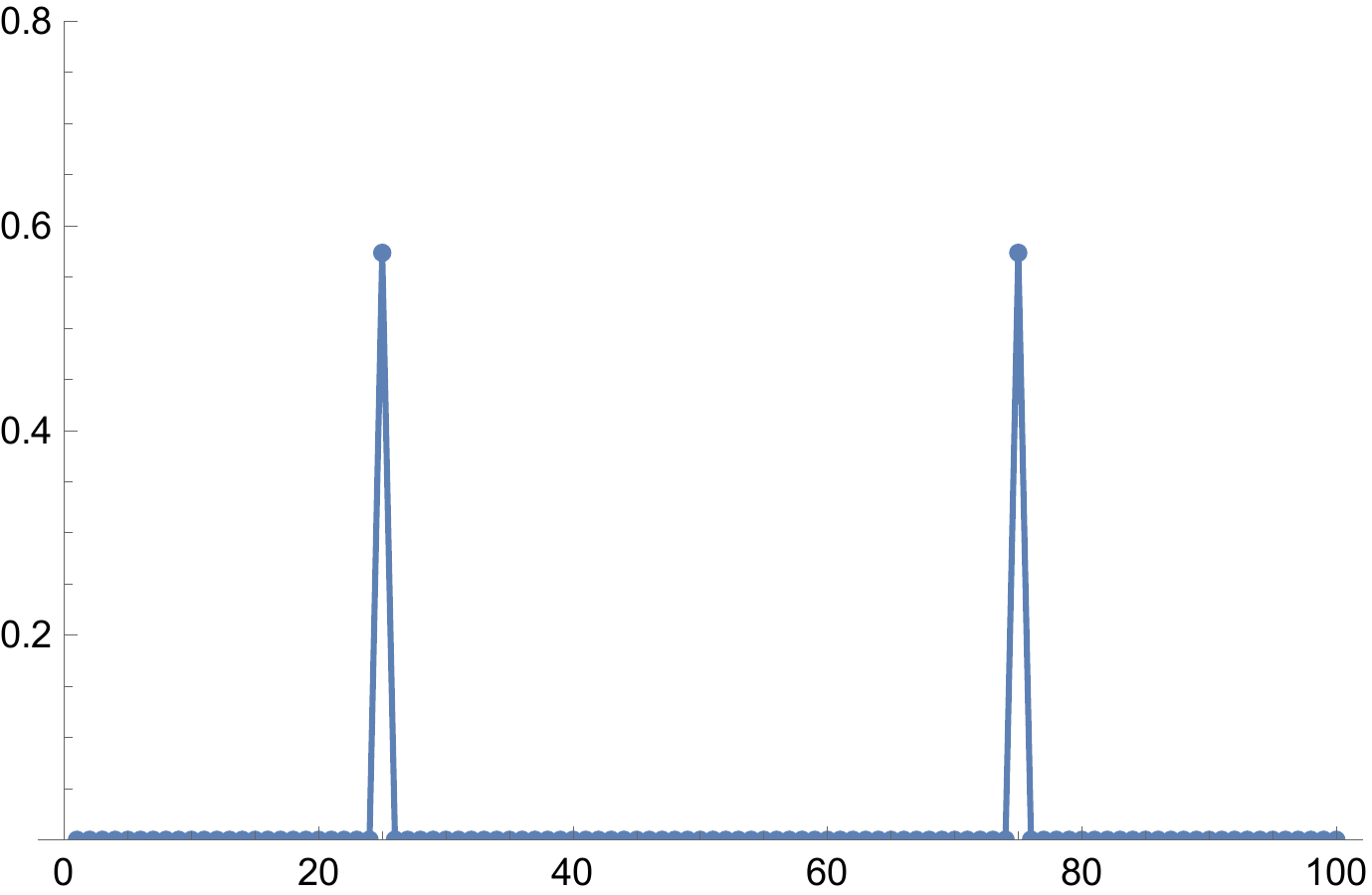}
    \caption{Plot of the overlap $\bra{\phi;a} {\cal O},\tau; a \rangle$
    for $-\pi<\tau <\pi$}
    \label{f3}
  \end{minipage}
  \begin{minipage}[b]{0.45\linewidth}
    \centering
    \includegraphics[width=7.5cm]{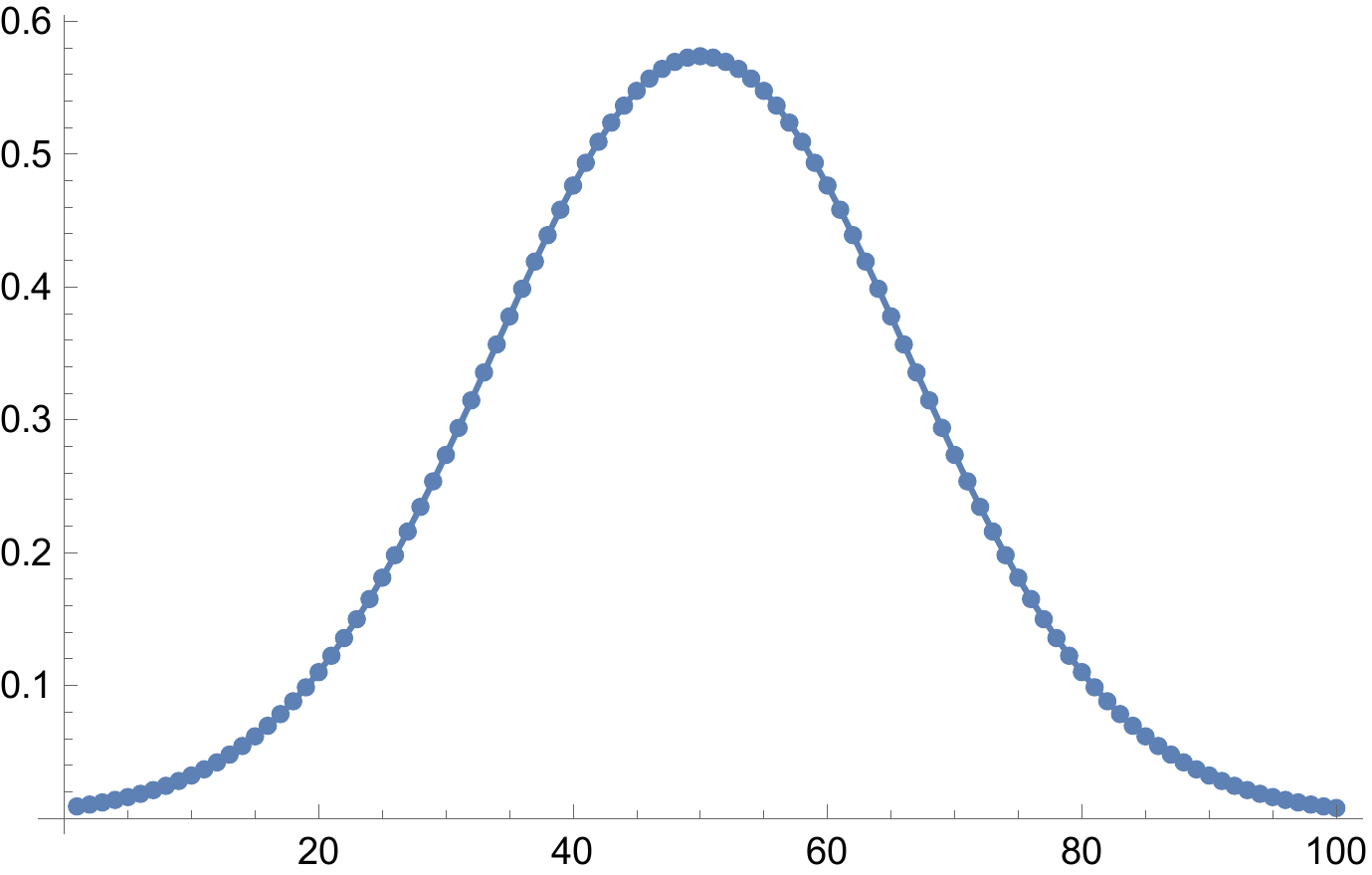}
    \caption{Plot of the overlap $\bra{\phi;a} {\cal O},\tau; a \rangle$ for $(\frac12 -\frac{1}{100} ) \pi  <\tau <(\frac12 +\frac{1}{100} ) \pi$}
    \label{f4}
  \end{minipage}
\end{figure}

We can numerically check this.
As an example, we will take
$d=3$, $\Delta=4.8$
and $a=0.005$.
The Figure \ref{f3} is the plot of the overlap $\bra{\phi;a} {\cal O},\tau; a \rangle$, where 
$\tau =\frac{2 \pi}{100} m -\frac{\pi}{2}$ and
the horizontal axis represents $m$.
This shows the sharp peaks 
at $\tau=\pi/2 + \pi \Z$.
The Figure \ref{f4} is the same plot of the overlap $\bra{\phi;a} {\cal O},\tau; a \rangle$, where 
$\tau =\frac{\pi}{2} +\frac{2 \pi}{10000} (m-50) $ and
the horizontal axis represents $m$.
This plot focuses on near $\tau=\pi/2$ region.

\begin{figure}[htbp]
  \begin{minipage}[b]{0.45\linewidth}
    \centering
    \includegraphics[width=7.5cm]{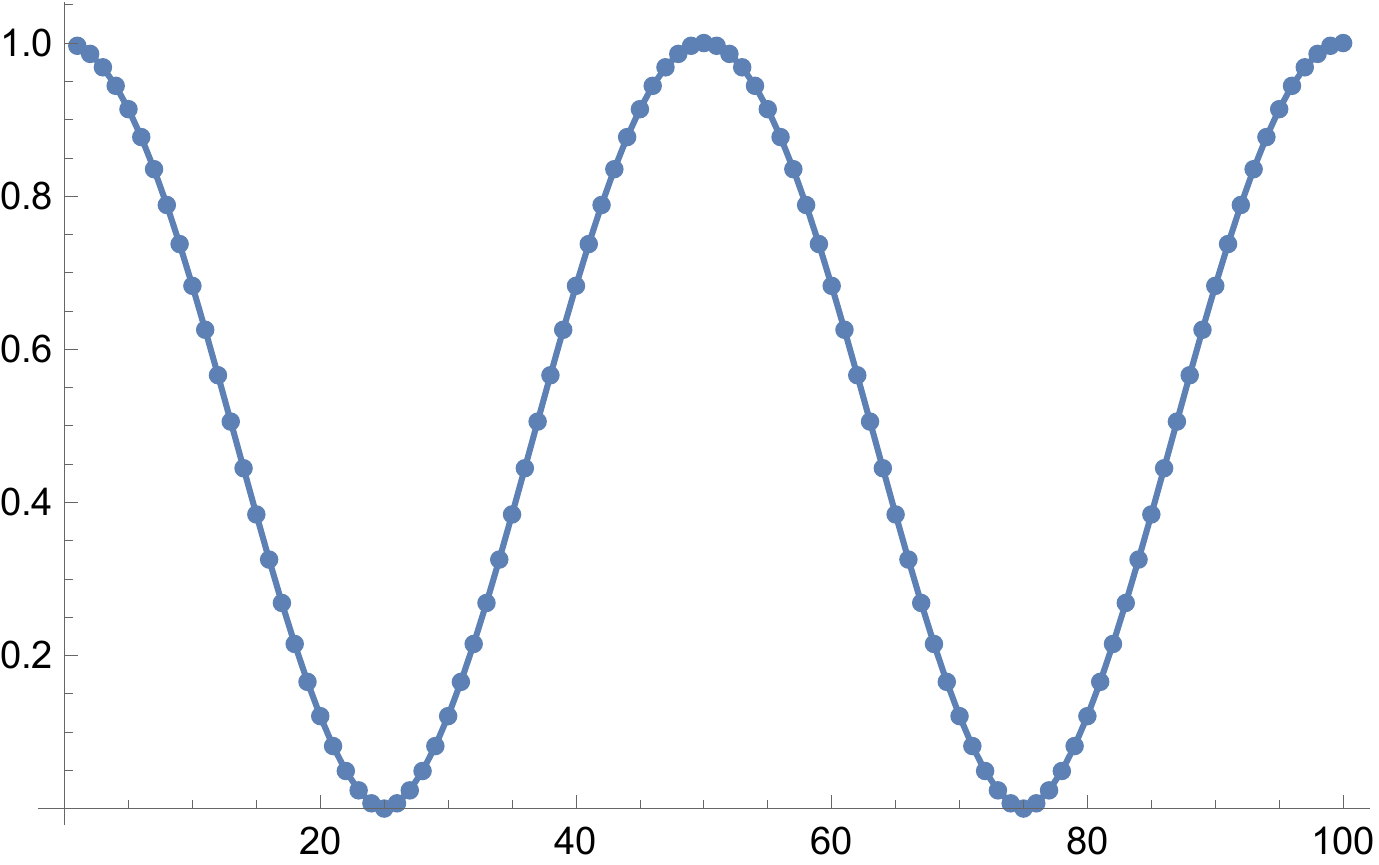}
    \caption{Plot of the HKLL smearing function ${1 \over | \cos \tau |^{d-\Delta} }$}
    \label{f5}
  \end{minipage}
  \begin{minipage}[b]{0.45\linewidth}
    \centering
    \includegraphics[width=7.5cm]{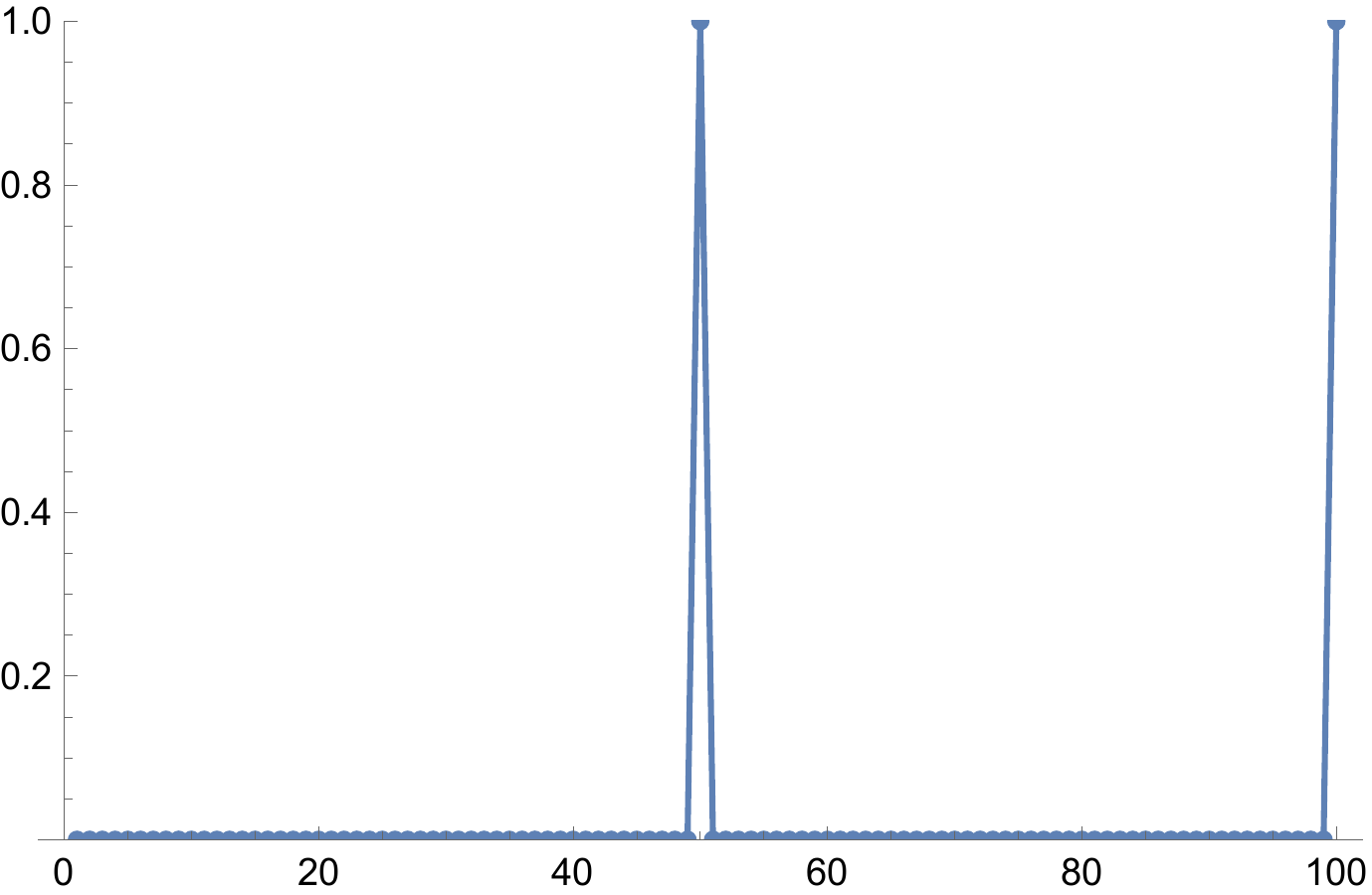}
    \caption{The corresponding two "particles" in the CFT picture}
    \label{f6}
  \end{minipage}
\end{figure}

For comparison, 
the Figure \ref{f5} is the plot of ${1 \over | \cos \tau |^{d-\Delta} }$, where 
$\tau =\frac{2 \pi}{100} m -\frac{\pi}{2}$ and
the horizontal axis represents $m$.
We also show the plot of
 $\bra{{\cal O},\tau;a} {\cal O},\tau'=0;a \rangle $
in Figure \ref{f6}, where 
$\tau =\frac{2 \pi}{100} m -\frac{\pi}{2}$ and
the horizontal axis represents $m$.
This clearly show the orthogonality.

\newpage 

\bibliographystyle{utphys}
\bibliography{main.bib}
\end{document}